# Enhancing Team Diversity with Generative AI: A Novel Project Management Framework

Johnny Chan
University of Auckland
jh.chan@auckland.ac.nz

Yuming Li
University of Auckland
yuming.li@auckland.ac.nz

***Abstract*— The proliferation of homogeneous teams in academic research projects limits the diversity necessary for innovative outcomes. This paper presents a framework that applies generative AI (GenAI) to project management, aiming to enhance team diversity by introducing GenAI agents modelled after a range of successful team member roles and personalities. These agents are designed to fill existing gaps in team composition, potentially leading to more dynamic and effective collaborations. Initial experiments with GenAI agents, fine-tuned with personality datasets, indicate a marked improvement in modelling and processing diverse personality traits. This framework promises to advance the practical application of AI in cultivating diverse and successful project teams, suggesting a significant shift in traditional project management methods towards more AI-integrated approaches. Future work will further explore the application of these GenAI agents in real-world settings, aiming to validate their effectiveness in fostering diversity and improving project outcomes.**

## I. INTRODUCTION

Projects are commonly defined as a series of coordinated activities to achieve specific objectives with clear start and end times, and typically distinct from routine operational activities. The essence of a project lies in its temporariness and creativity, meaning they are established to produce unique products, services, or results on a one-off basis [1]. Projects can be categorised into various types depending on their nature, objectives, and areas of application, such as construction projects, organisational projects, scientific research projects, and social arts projects [2]. Scientific research projects, especially those within academic institutions, usually comprise a series of systematic activities that explore, validate, or develop specific scientific questions or hypotheses through scientific methods. These projects often revolve around the cycles of research funding. Compared to traditional, result-oriented corporate projects, scientific research projects focus more on theoretical innovation and ground-breaking optimisation, aiming to set long-term potential goals to push the boundaries of knowledge. These projects are often the breeding grounds for new technologies, theories, or even new industries, and they have immeasurable potential contributions to societal progress and the knowledge economy.

However, despite the significant role that scientific research projects play in driving academic and technological innovation, they face several challenges, particularly in the process of commercialisation due to their breeding environment. Firstly, the composition of scientific research project teams is often homogeneous, mostly consisting of students and academic staffs. These members are likely to share similar educational backgrounds and theoretical knowledge, but lack the diversity of social experience or even personality found in the corporate environments. This homogeneity may limit the team's ability to solve complex problems and turn them into successful business implementation. Secondly, the ideal project team usually needs to include different roles, such as operators, leaders, engineers, developers, etc, each of whom contributes indispensably to the project's success. A specific project role and teammate personality composition pattern have been proven to enhance the rate of successful commercialisation of a project [3]. However, in student-led scientific research projects, it is often difficult to achieve such role diversity. Due to funding limitation and the instability nature of some research projects, attracting teammates to join from outside the university could be difficult. The recruitment of members for on-campus research projects often lacks professional human resources management, further restricting the team's capabilities in hiring, assessment, training, and professional development. These challenges ultimately could lead to sub-optimal progress and outcome, no matter how well the project is managed from all other dimensions.

With the generative AI (GenAI) wave initiated by ChatGPT in late 2022 [4], GenAI may offer new possibilities for project management. In this paper, we propose a GenAI-empowered project management framework, utilising GenAI agents to simulate team members playing different roles. This approach is theoretically based on insights from sociological research, which underscores the interplay between project team diversity, founder personality, and entrepreneurial success [5], [6], [7]. Building on this foundation, we suggest the integration of GenAI agents in original project team. Our GenAI agents are designed to adopt a range of personalities and roles tailored to university-led project teams, thereby composing and structuring teams that mirror the personality compositions identified in successful entrepreneurial patterns.

## II. RELATED WORK

Project management, as a practice of organising and controlling resources, aims to effectively guide a project from conception to completion. Traditional project management processes usually follow a standardised phase division, including initiation, planning, execution, monitoring and controlling, and closing [8]. Although this process provides a clear framework for project management, it also faces challenges in dealing with the increasing complexity and dynamism of projects. For instance, as project scales enlarge and team distributions become more widespread, coordination and communication become more challenging, and traditional methods may lack flexibility and adaptability [9].

To address the limitations, AI technology may bring in new possibilities to the field of project management. AI can assist in decision-making by analysing historical project data, predicting risks, optimising resource allocation, and even automating certain management tasks [10]. Castañé et al. [11] combine human control with responsible AI to facilitate functions in manufacturing project management, including process planning, production planning, scheduling, and real-time control. Their goal is to achieve optimal complementarity between humans and AI systems. GenAI is an emerging branch within the AI domain, capable of not only analysing and processing data but also generating new content and solution. The application of GenAI in project management spans multiple aspects. For instance, GenAI can be utilised for risk assessment and management, predicting and mitigating potential risks through the analysis of historical data [12]. Furthermore, GenAI can be deployed to optimise resource allocation, automate project reporting, and even simulate interactions and collaborations among project team members [13].

However, existing applications of GenAI in project management, whether in research or mature products, primarily focus on data analysis, predicting resource allocation. While these applications play a significant role in enhancing efficiency and reducing human errors, they largely concentrate on quantitative data processing and tend to overlook the complex diversity and psychological factors of the human element, which is central to project management. Our proposed new project management framework, which utilises GenAI technology to simulate missing roles and personality traits in project teams, expands the application scope of GenAI. It offers a novel perspective, extending the capabilities of AI from purely data processing to a deeper understanding and enhancement of human team collaboration, thereby improving the efficacy of project management.

## III. METHODOLOGY

In this paper, we present a project management framework empowered by GenAI agents. This framework adheres to the basic process of traditional project management (initiation, planning, execution, monitoring and controlling, and closing), while integrating an innovative module - team analysis. The framework is shown in Fig. 1 During the initiation phase, the fundamental objectives and scope of the project are defined. Before the traditional planning stage, the inserted team analysis module classifies team members' personalities and roles, and then, following the ideal team member composition pattern [3], identifies gaps and personality diversity needs within the team. GenAI agents, at this stage, are fine-tuned with a personality dataset on ChatGPT to learn and simulate the traits of each team role and personality. Subsequently, existing team members and simulated agent members continue with the traditional project management process, developing detailed project plan during the planning phase, including timeline, budget, resource allocation, and risk management strategy. In the execution phase, team members are assigned tasks, resources are allocated, and project activities are carried out according to the plan. The monitoring and controlling phase involves continuous monitoring and assessment of project progress to ensure alignment with the plan. Finally, key tasks of the closing phase include closing all project activities, ensuring all contractual terms are met, formally accepting the final product or service, and conducting project summarisation.

Given that the focus of this paper is on the technical innovation and the preliminary application of GenAI in project management, this section emphasises and validates our proposed GenAI agent and its feasibility.

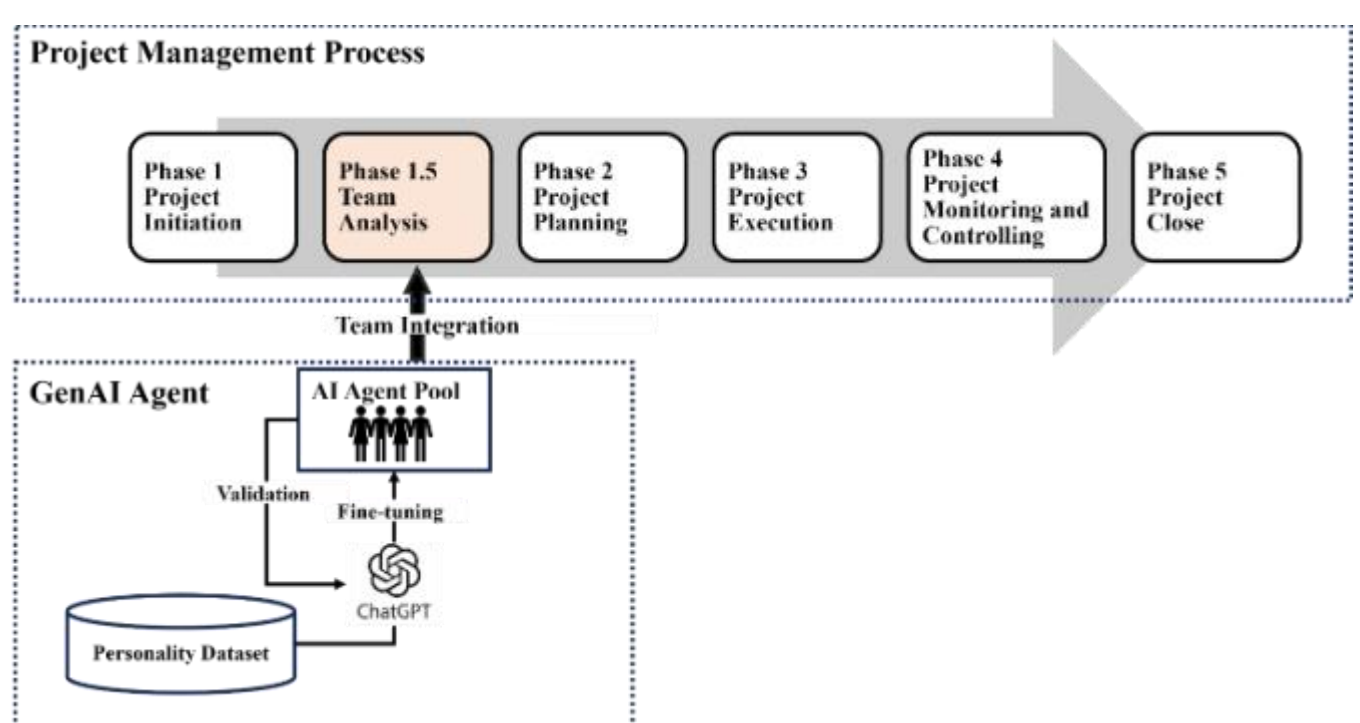

Fig. 1. The project management framework empowered by GenAI agent

### A. *GenAI Agent*

GenAI is a cutting-edge application of AI; unlike discriminative models that focus on classifying or predicting outcomes from input data, generative models aim to create entirely new outputs that are statistically similar to the training data [14]. This capability enables GenAI to excel in the creation of diverse types of data, such as images, text, and audio. Key technologies underpinning GenAI include Generative Adversarial Networks (GANs), which involve a competitive game between networks during training, and Transformer models, which efficiently process sequential data and capture long-distance dependencies in text through the Self-Attention mechanism [15]. As a revolutionary text generation model based on GenAI technology, ChatGPT, trained on extensive textual data, is capable of generating coherent, relevant, and contextually appropriate textual responses. Its training encompasses two phases: pre-training and fine-tuning. During the pre-training stage, the model is trained on a broad dataset of textual information to master the fundamental rules and structure of language. The fine-tuning stage is then tailored for specific application scenario [16]. In this paper, we select *'gpt-3.5-turbo-1106'* as the base model for our GenAI agent and conduct fine-tuning on this basis to incorporate personality traits.

Fine-tuning refers to the process of optimising a pre-trained model for specific tasks. In this paper, our fine-tuning objective is to equip ChatGPT with the cognition of personalities and designated roles. Although the ChatGPT API includes a 'role' attribute, it focuses more on specific identity played in certain context, such as assistant, system or user. These roles are designed to cater specific interaction scenario, but they are insufficient to simulate or understand insightful personality traits. In our experiments elaborated in the following section, we compared the interpretation and understanding of personality information by the ChatGPT model without fine-tuning, thereby confirming the necessity of fine-tuning on personality dataset.

### B. Personality Dataset

In terms of defining personality, we base our categorisation of personality traits in the text on the Big Five personality model [17]. The Big Five personality traits refers to agreeableness conscientiousness, extraversion, openness and neuroticism. It provides a systematic method to describe and predict individual behaviours and preferences. For our fine-tuning dataset, we chose FriendsPersona [18], the first dialogue-based personality dataset, which contains dialogues based on the characters from the TV show "Friends" and their corresponding multi-label annotations of the Big Five personality traits. The data distribution is illustrated in Table 1, where Pos. means the entry contains corresponding personality traits, Neg. means the entry does not contain corresponding personality traits.

TABLE I. FRIENDSPERSONA DATASET DESCRIPTION

| Personality | Description | Neg. | Pos. |
|---|---|---|---|
| Agreeableness | Tendency towards cooperation and social harmony, with compassionate, cooperative, empathetic, trusting, helpful characteristics. | 306 | 405 |
| Conscientiousness | Tendency towards organization and planning, with self-discipline, organised, reliable, cautious, hardworking characteristics. | 381 | 330 |
| Extraversion | Outward orientation towards social world, with sociability, assertiveness, high energy, positive emotions, expressiveness characteristics. | 399 | 312 |
| Openness | Openness to new experiences, with curiosity, creativity, sensitivity to art and beauty, willingness to try new things characteristics. | 249 | 462 |
| Neuroticism | Tendency to experience negative emotions, with anxiety, sadness, moodiness, emotional instability, prone to stress characteristics. | 379 | 332 |

As shown in Table 1, the dataset has a moderate number of samples with a relatively balanced distribution. Subsequently, we pass the raw data to data pre-processing. Our data processing includes data cleaning and format conversion. As illustrated in Fig 2, given that the raw data was extracted directly from the "Friends" series using the MainSpeakerFinder algorithm [18] for identifying main speakers and their lines, the extracted data contains HTML symbols and character names is irrelevant to fine-tuning, such as "***<b>s01_e01_c02(0) for Chandler Bing</b><br><br><b>Monica Geller</b>:*** ". Therefore, we first cleanse the raw data to retain only the actual dialogue content. Considering that when fine-tuning with the ChatGPT API, data is typically formatted in JSONLines, with each line being a complete JSON object. The fine-tuning data needs to be organised as a series of interactive message objects, each containing the roles in the dialogue (i.e. assistant, system or user) and their corresponding content. 'Assistant' represents the chatbot itself, responding to user input, with its content being the answers to the potential prompts for the ChatGPT model to learn. 'System' refers to the entity that sets context or provides guiding information in the dialogue. 'User' refers to the human user interacting with the chatbot, whose content provides potential prompts for the ChatGPT model to learn. An example of our data pre-processing is shown in Figure 2:

```
<b>s01_e01_c02(0) for Chandler Bing</b><br><br><b>Monica Geller</b>: Now I'm guessing that he bought her the big pipe
organ, and she's really not happy about it.<br><br><b>Chandler Bing</b>: (imitating the characters) Tuna or egg salad?
Decide!<br><br><b>Ross Geller</b>: (in a deep voice) I'll have whatever Christine is having.<br><br><b>Rachel Green</b>:
(on phone) Daddy, I just... I can't marry him! I'm sorry. I just don't love him. Well, it matters to me!<br><br>(The scene on TV has
changed to show two women, one is holding her hair.)
```

*Pre-processing*

```
: Now I'm guessing that he bought her the big pipe organ, and she's really not happy about it.
: (imitating the characters) Tuna or egg salad? Decide!
: (in a deep voice) I'll have whatever Christine is having
: (on phone) Daddy, I just... I can't marry him! I'm sorry. I just don't love him. Well, it matters to me!<br><br>(The scene on TV has
changed to show two women, one is holding her hair.)
```

*Formatting*

```
{"messages": [{"role": "system", "content": "you are a person with agreeableness personality."}, {"role": "user", "content":
"Now I'm guessing that he bought her the big pipe organ, and she's really not happy about it."}, {"role": "assistant", "content":
"(imitating the characters) Tuna or egg salad? Decide!"}]}
{"messages": [{"role": "system", "content": "you are a person with agreeableness personality."}, {"role": "user", "content":
"(imitating the characters) Tuna or egg salad? Decide!"}, {"role": "assistant", "content": "(in a deep voice) I'll have whatever
Christine is having."}]}
{"messages": [{"role": "system", "content": "you are a person with agreeableness personality."}, {"role": "user", "content": "(in
a deep voice) I'll have whatever Christine is having."}, {"role": "assistant", "content": "(on phone) Daddy, I just... I can't marry
him! I'm sorry. I just don't love him. Well, it matters to me!(The scene on TV has changed to show two women, one is holding
her hair.)"}]}
```

Fig. 2. The sample of data pre-processing

Finally, the pre-processed dataset was divided into a training set and a validation set by an 8:2 ratio, with the training set serving as the primary dataset for fine-tuning the model. The validation set was used to assess model performance and adjust hyperparameters during the fine-tuning process. The aim of the aforementioned fine-tuning experiment was to impart an understanding of personality traits to the GenAI agent. Subsequently, we conducted tests on Essay dataset [19] annotated with the same Big Five personality traits to demonstrate the effectiveness of the fine-tuned model and the feasibility of the agent assuming the roles of different personality team members. The final results and analysis are elucidated in the next section.

## IV. RESULTS AND ANALYSIS

In accordance with the experimental setup outlined in the previous section, we trained the 'gpt-3.5-turbo-1106' model on the FriendsPersona dataset, and the training accuracy and loss are depicted in Fig 3. This figure illustrates the progression of training accuracy for five different personality categories: Agreeableness (AGR), Conscientiousness (CON), Extraversion (EXT), Openness (OPN), and Neuroticism (NEU) over the course of the training period. Each category is represented by a distinct colour, which enhances the clarity of comparison. From Fig 3, we can infer that while the general trends of training accuracy and loss confirm the effectiveness of the model's learning process and indicate the absence of overfitting, there is still some noticeable fluctuation. This suggests the need for further investigation into the model's training or data quality.

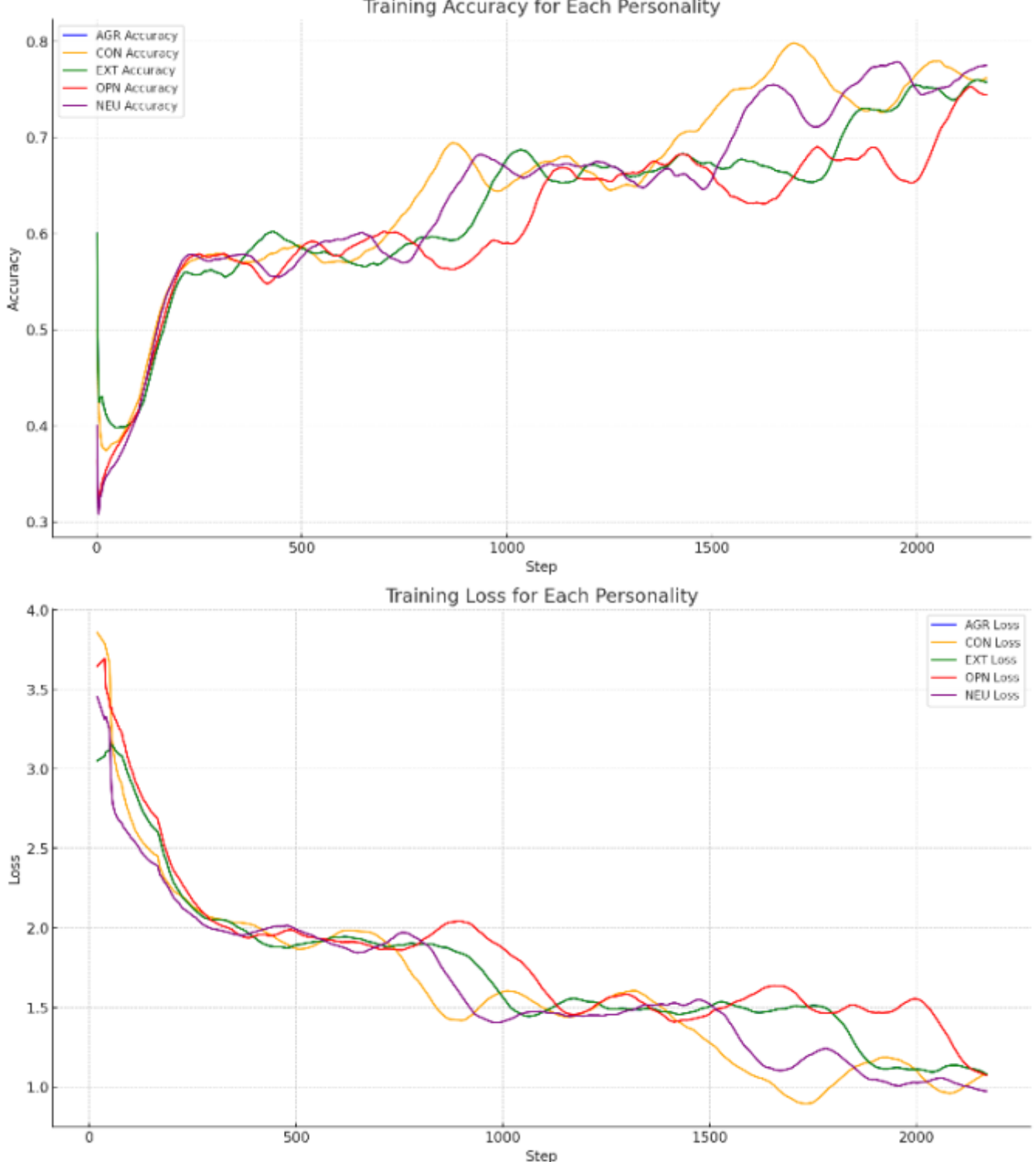

Fig. 3. The training accuracy and training loss for each personality

Following this, we will conduct testing on the fine-tuned model using the Essay dataset [19]. The Essay dataset is a substantial compilation of text generated from the stream of consciousness. This dataset consists of contributions from 2,467 users who were categorised based on the Big Five personality traits, aligning with the labelling format of FriendsPersona described in the previous section. Each essay in this dataset is accompanied by a label indicating the author's personality trait, making it well-suited for supervised learning tasks. The statistics of Essay dataset is listed in Table 2. However, due to privacy concerns and incompatibility with ChatGPT's terms of use, we are unable to employ this dataset for the purpose of fine-tuning. Instead, we have opted to use it as a testing dataset to evaluate the efficacy of our fine-tuning efforts on the FriendsPersona dataset.

TABLE II. ESSAY DATASET DESCRIPTION

| **Personality** | **Neg.** | **Pos.** |
|---|---|---|
| Agreeableness | 1157 | 1157 |
| Conscientiousness | 1214 | 1214 |
| Extraversion | 1191 | 1191 |
| Openness | 1196 | 1196 |
| Neuroticism | 1234 | 1234 |

From Table 2, it is evident that the distribution of the Essay dataset is more balanced, ensuring consistency between positive and negative samples. Subsequently, we conducted comparative experiments on this dataset. We used the original 'gpt-3.5-turbo-1106' model as our baseline and a fine-tuned model as the comparison group. For each entry in the test set, we performed binary classification of personality traits by calling the ChatGPT API. For example, when assessing the openness attribute of an entry, the prompt used in calling the API was:

*"Prompt: I require an analysis to classify the following text under one of the Big Five personality traits (Agreeableness, Conscientiousness, Extraversion, Openness, Neuroticism). The question is: Does the provided text demonstrate or suggest personality associated with the Openness trait? Please respond exclusively with 'Yes' or 'No'. Text: {text}"*

The final test results, as shown in Table 3, utilise a metric system comprising Precision, Recall, and F1 Score. 'Precision' measures the accuracy of positive predictions, 'Recall' evaluates the model's capability to identify all pertinent cases, and the 'F1 Score' is the harmonic mean of precision and recall, offering a balanced perspective of both metrics.

TABLE III. TEST RESULT ON ESSAY DATASET

| **Personality** | **Model** | **Precision** | **Recall** | **F1 Score** |
|---|---|---|---|---|
| Agreeableness | baseline | 0.597 | 0.187 | 0.285 |
| | Fine-tuned | 0.574 | 0.514 | 0.542 |
| Conscientiousness | baseline | 0.482 | 0.157 | 0.237 |
| | Fine-tuned | 0.505 | 0.596 | 0.547 |
| Extraversion | baseline | 0.547 | 0.266 | 0.358 |
| | Fine-tuned | 0.558 | 0.201 | 0.296 |
| Openness | baseline | 0.553 | 0.363 | 0.439 |
| | Fine-tuned | 0.564 | 0.511 | 0.536 |
| Neuroticism | baseline | 0.569 | 0.614 | 0.590 |
| | Fine-tuned | 0.530 | 0.760 | 0.625 |

From Table 3, it can be discerned that fine-tuning has an overall positive impact on the recognition performance for most personality categories. For instance, the F1 scores improved for all categories except 'Extraversion' after fine-tuning, albeit at the cost of precision in some cases. This indicates that fine-tuning effectively balances the model's ability to identify relevant instances and reduce false positives. This preliminary evidence supports the hypothesis that GenAI agents, trained on personality datasets, can better understand and recognise personalities. It also validates the feasibility of the proposal to train virtual teammates with diverse personalities and roles using GenAI technology, thereby providing momentum for the next stage of Gen agent development.

## V. Conlusion and Future Work

In this paper, we propose a GenAI empowered project management. This framework integrates patterns of team member personalities and roles that are more likely to succeed according to sociological theories. It employs AI agents to fill gaps in team personality or roles, aiming to address challenges in project advancement and implementation, particularly in homogenous personnel compositions within university or research institution project teams. The framework enhances traditional project management processes by incorporating analysis of team member personalities and roles, filling specific team vacancies with GenAI agents fine-tuned on a personality dataset. Our preliminary experiments demonstrate significant improvements in the model's understanding and handling of personality traits after fine-tuning with a personality dataset, thereby validating the feasibility of GenAI teammates.

In future work, we intend to extend the fine-tuning of GenAI teammates beyond personality traits to include professional roles such as operator, leader, engineer, and developer. We plan to employ synthetic datasets for this purpose, with a focus on validating the synthetic data quality. Additionally, we aim to compare the impact of GenAI agents versus non-GenAI agents on project progress, output, and commercialisation within real-world projects.